\documentclass[smus]{snow2e}
\usepackage{psfig}

\def\nn{\noindent}
\def\Re{{\cal R \mskip-4mu \lower.1ex \hbox{\it e}\,}}
\def\Im{{\cal I \mskip-5mu \lower.1ex \hbox{\it m}\,}}
\def\ie{{\it i.e.}}

\def\etal{{\it et al.}}
\def\ibid{{\it ibid}.}
\def\sub#1{_{\lower.25ex\hbox{$\scriptstyle#1$}}}

\def\to{\rightarrow}

\def\subw{_{\rm w}}
\def\mh{\ifmmode m\sbl H \else $m\sbl H$\fi}
\def\mch{\ifmmode m_{H^\pm} \else $m_{H^\pm}$\fi}
\def\mt{\ifmmode m_t\else $m_t$\fi}
\def\mc{\ifmmode m_c\else $m_c$\fi}
\def\mz{\ifmmode M_Z\else $M_Z$\fi}
\def\mw{\ifmmode M_W\else $M_W$\fi}
\def\mws{\ifmmode M_W^2 \else $M_W^2$\fi}
\def\mhs{\ifmmode m_H^2 \else $m_H^2$\fi}   
\def\mzs{\ifmmode M_Z^2 \else $M_Z^2$\fi}
\def\mts{\ifmmode m_t^2 \else $m_t^2$\fi}
\def\mcs{\ifmmode m_c^2 \else $m_c^2$\fi}
\def\mchs{\ifmmode m_{H^\pm}^2 \else $m_{H^\pm}^2$\fi}
\def\ztwo{\ifmmode Z_2\else $Z_2$\fi}
\def\zone{\ifmmode Z_1\else $Z_1$\fi}
\def\mtwo{\ifmmode M_2\else $M_2$\fi}
\def\mone{\ifmmode M_1\else $M_1$\fi}
\def\tb{\ifmmode \tan\beta \else $\tan\beta$\fi}
\def\xw{\ifmmode x\subw\else $x\subw$\fi}
\def\ch{\ifmmode H^\pm \else $H^\pm$\fi}
\def\lum{\ifmmode {\cal L}\else ${\cal L}$\fi}
\def\inpb{\ifmmode {\rm pb}^{-1}\else ${\rm pb}^{-1}$\fi}
\def\infb{\ifmmode {\rm fb}^{-1}\else ${\rm fb}^{-1}$\fi}
\def\epem{\ifmmode e^+e^-\else $e^+e^-$\fi}
\def\ppb{\ifmmode \bar pp\else $\bar pp$\fi}
\def\bsg{\ifmmode B\to X_s\gamma\else $B\to X_s\gamma$\fi}
\def\bsll{\ifmmode B\to X_s\ell^+\ell^-\else $B\to X_s\ell^+\ell^-$\fi}
\def\bstt{\ifmmode B\to X_s\tau^+\tau^-\else $B\to X_s\tau^+\tau^-$\fi}
\def\nng{\ifmmode e^+e^-\to\nu\bar\nu\gamma\else $e^+e^-\to\nu\bar\nu\gamma$\fi}

\newskip\zatskip \zatskip=0pt plus0pt minus0pt
\def\matth{\mathsurround=0pt}
\def\lsim{\mathrel{\mathpalette\atversim<}}
\def\gsim{\mathrel{\mathpalette\atversim>}}
\def\atversim#1#2{\lower0.7ex\vbox{\baselineskip\zatskip\lineskip\zatskip
  \lineskiplimit 0pt\ialign{$\matth#1\hfil##\hfil$\crcr#2\crcr\sim\crcr}}}

\begin{document}
\onecolumn

\vbox{ \large
\begin{flushright}
SLAC-PUB-7441\\
\end{flushright}

\vspace{1in}

\begin{center}
{\Large\bf Signatures of Extended Gauge Sectors in $e^+e^-\to 
\nu\bar\nu\gamma$$^*$}
\medskip
\vskip .3cm

{\large JoAnne L. Hewett }\\
Stanford Linear Accelerator Center, Stanford, CA 94309, USA\\
\vskip 1.0cm

\vskip3cm

{\bf Abstract}\\
\end{center}

The ability of high energy \epem\ colliders to indirectly probe the existence 
of heavy new charged gauge bosons via their exchange in the reaction \nng\ 
is investigated.  It is shown that examination the resulting photon energy 
spectrum with polarized beams can extend the $W'$ search reach above the 
center of mass energy. 

\vspace*{2.5in}
\noindent To appear in the {\it Proceedings of the 1996 DPF/DPB Summer
Study on New Directions for High Energy Physics - Snowmass96}, Snowmass, CO,
25 June - 12 July 1996.
\vskip 1.3in

\noindent $^*$Work supported by the U.S. Department of Energy under contract
DE-AC03-76SF00515. 

\thispagestyle{empty}
}
\newpage

\twocolumn
\title{Signatures of Extended Gauge Sectors in $e^+e^-\to \nu\bar\nu\gamma$
\thanks{Work supported by the U.S. Department of Energy under contract
DE-AC03-76SF00515. }}

\author{JoAnne L. Hewett \\
{\em Stanford Linear Accelerator Center, Stanford, CA 94309, USA}\\
}

\maketitle

\begin{abstract}
The ability of high energy \epem\ colliders to indirectly probe the existence 
of heavy new charged gauge bosons via their exchange in the reaction \nng\ 
is investigated.  It is shown that examination the resulting photon energy 
spectrum with polarized beams can extend the $W'$ search reach above the 
center of mass energy. 
\end{abstract}
\vspace*{5mm}

Extended gauge sectors are a feature in many scenarios of physics beyond
the Standard Model (SM).  Perhaps the most appealing set of
enlarged electroweak models are those which arise in supersymmetric
grand unified theories (GUTs), such as SO(10) and $E_6$.  However, more complex
gauge structures are also present in several non-GUT scenarios, including 
models of composite gauge bosons, horizontal interactions, and topcolor 
assisted technicolor.  The existence of new gauge bosons is the hallmark 
of such theories and the discovery of such particles would provide 
uncontested evidence for new physics with an extended gauge sector, and the 
study of their couplings would provide a diagnostic tool in determining the 
underlying theory.

The conventional search strategy for new gauge bosons at hadron colliders is 
through their direct production and subsequent decay to lepton pairs, \ie, the 
Drell-Yan mechanism.  Present bounds\cite{tev} on the mass of a new neutral 
gauge boson, $Z'$, from the Tevatron are generally in the range $M_{Z_2}\gsim
500-700$ GeV, with the exact value being dependent on the specific extended 
model.  This search reach is expected\cite{future} to increase by $\sim 300$ 
GeV with $1 \infb$ of luminosity.  The $Z'$ discovery potential at the LHC 
is in the mass range of $4-5$ TeV.  The corresponding Tevatron bound\cite{d0wp}
on the mass of a new charged boson, $W'$, is 720 GeV; this limit
assumes that the $W'$ has SM strength couplings and that it decays into 
a light and stable neutrino which manifests itself in the detector as missing 
$E_T$.  The LHC discovery reach\cite{future}  for a heavy charged boson, with 
the same assumptions, is $\sim 5.9$ TeV  with 100 \infb.  These bounds are 
invalidated if the $W'$ decays into a heavy  neutrino\cite{d0wp}, or as 
discussed below,  can be seriously degraded for some regions of parameter space 
in specific extended electroweak models.

At \epem\ colliders direct production of new gauge bosons is kinematically
limited by the available center of mass energy, and hence different search
tactics are necessary.  Such machines, however, easily allow for $Z'$ indirect 
searches for the case $M_{Z_2}>\sqrt s\,$ by looking for deviations from SM 
expectations for cross sections and asymmetries associated with fermion pair 
production\cite{future,jlhtgr}.  This is similar to the exploration of the 
modifications of QED predictions due to the SM $Z$ boson at PEP/PETRA.  The 
bounds obtained in this manner\cite{future} are greatly model dependent, but
lie in the general range of $2-5$ TeV for $\sqrt s=500$ GeV with 50 \infb.  
It has been believed that $W'$ searches may only proceed via
direct production,  with the most promising process, $\epem\to W'W'^*\to W'jj$, 
resulting in a discovery reach\cite{tgr1} of $M_{W'}\simeq 0.8\sqrt s$.  
Indirect $W'$ signals are thought to be inaccessible as charged gauge bosons
do not contribute to $\epem\to f\bar f$ ($f\neq\nu$).  

Here we explore the neutrino counting reaction, \nng, to determine if 
the $W'$ search reach can be extended to masses above $\sqrt s$.  
In the SM, this process proceeds through s-channel $Z$ and t-channel $W$ 
exchange with the photon being radiated from every possible charged particle.  
Although the resulting cross section suffers from an additional factor of 
$\alpha$ and from 3-body phase space suppression 
(as compared to fermion pair production) , it still has a reasonable
value\cite{nlcstudy}, of order a few pb at a 500 GeV NLC.  In extended gauge 
models it is modified by both s-channel $Z'$ and t-channel $W'$ exchange.
The influence of additional $Z'$ exchange alone has been previously 
examined\cite{nngam} in $E_6$ GUTs models at SLC/LEP energies, where the effects
were found to be small due to the close proximity of the SM $Z$ resonance.

We examine the effect of two extended electroweak models on this reaction.
The first is the Left-Right Symmetric Model (LRM)\cite{lrm}, based on
the gauge group $SU(3)_C\times SU(2)_L\times SU(2)_R\times U(1)_{B-L}$, which  
has right-handed charged currents, and hence restores parity at a higher mass
scale.  This model contains the free parameter $\kappa\equiv g_R/g_L$, which
represents the ratio of the right- to left-handed coupling strengths and lies
in the range $0.55\lsim\kappa\lsim 2.0$.  Strict left-right symmetry dictates
that $\kappa=1$.  It has been shown\cite{desh} that
a light right-handed mass scale ($M_R\sim 1$ TeV) is consistent with coupling
constant unification in supersymmetric SO(10) GUTs.  A direct mass relationship
between the right-handed charged and neutral gauge bosons is present and is
given by
\begin{equation}
{M^2_{Z_R}\over M^2_{W_R}}={\kappa^2(1-x_w)\rho_R\over \kappa^2(1-x_w)-x_w}\,,
\end{equation}
where $\rho_R=1 (2)$ probes the symmetry breaking of the $SU(2)_R$
by right-handed Higgs doublets (triplets) and $x_w=\sin^2\theta_w$.  The 
fermionic $Z_R$ couplings are fixed and can be written as 
$(g/2c_w)(\kappa-(1+\kappa)x_w)^{-1/2}[x_wT_{3L}+\kappa(1-x_w)T_{3R}
-x_wQ]$, with $T_{3L(R)}$ being the fermion's left-(right-) handed isospin, and
$Q$ is the fermion electric charge.  In this study, we assume that the 
neutrinos are light and Dirac in nature.  We note the existence 
of a right-handed Cabbibo-Kobayashi-Maskawa (CKM) matrix in
this model, which need not be the same as the corresponding left-handed mixing
matrix.  This can degrade the $W_R$ search capability in hadronic 
collisions\cite{tgr2}, as
the weights of the various parton densities which enter the production cross
section can be dramatically altered.  For example, this could reduce the current
Tevatron $W_R$ mass bounds by up to a factor of $\sim 2$.

The second model we consider is the un-unified model (UUM)\cite{uum}, which
is based on the gauge group $SU(2)_q\times SU(2)_\ell\times U(1)_Y$, where
the quarks and leptons transform under their own $SU(2)$.  In this case
the additional heavy $W$ and $Z$ are approximately degenerate, $M_{W_H}
\simeq M_{Z_H}$.  The $Z_H$ fermionic couplings take the form $c_w[T_{3q}/\tan
\phi-\tan\phi T_{3\ell}]$, where $T_{3q(\ell)}$ is the $SU(2)_{q(\ell)}$ third
component of isospin, and $\phi$ is a mixing parameter which is constrained
to the range $0.24\lsim\sin\phi\lsim 0.99$.  Note that the additional neutral 
and charged currents are purely left-handed in this case.

In this preliminary study we employ the point interaction approximation for
the $W_i$ boson exchange diagrams in \nng.  This approximation is reasonable for
the heavy $W'$ contribution as we are considering the case $M_{W'}>
\sqrt s$, but it is known\cite{approx} to break down for the SM $W$ for center 
of mass energies above the $Z$ pole.  An exact calculation will be presented 
elsewhere\cite{jlh}.  In this approximation
the \nng\ differential cross section can then be cast in the form\cite{nngam}
(after integration over the angle of the final state photon)
\begin{equation}
{d\sigma\over dx} = {\alpha\over\pi x}\left[ (2-2x+x^2)\ln{1+\delta\over
1-\delta} -x^2\delta\right] \sigma^{\nu\bar\nu}_M(x)\,,
\end{equation}
where $x=2E_\gamma/\sqrt s$ and $\delta=\cos\theta_{\rm min}$ with 
$\theta_{\rm min}$ being the minimum angle between the initial electron beam
and the outgoing $\gamma$ allowed by the experimental cuts.  In our analysis
we take $\theta_{\rm min}=20^\circ$ and $E_\gamma>0.1E_{\rm beam}$.  The
angular cut corresponds to a conservative estimate of the acceptance of
a typical NLC detector\cite{nlcstudy} and is effective in removing background
from the process $\epem\to\epem\gamma$, while the minimum photon energy
ensures the finiteness of the cross section by removing the infrared and
collinear divergences.  $\sigma^{\nu\bar\nu}_M(x)$ is the cross section for 
the subprocess $\epem\to\nu\bar\nu$ evaluated at the center of mass energy 
$M^2=s(1-x)$.  

We now evaluate the $\epem\to\nu\bar\nu$ subprocess cross sections for 
polarized beams in our two extended electroweak models.
In the LRM these are given by
\begin{eqnarray}
\sigma_L(x) & = & {G_F^2M^2\over 6\pi} \left[ 1 -2M^2_{Z_1}\sum_i
{C_i(v_i^L+a_i^L)F_i} \right. \nonumber\\
& + & \left.  2N_\nu M^4_{Z_1}\sum_{i,j}(v^L_iv^L_j+a^L_ia^L_j)
(C_iC_j+\tilde C_i\tilde C_j)P_{ij}\right] \nonumber \\
\sigma_R(x) & = & {G_F^2M^2\over 6\pi}\left[ \left({\kappa M_L\over M_R}
\right)^4\right.\\
& - &\left. \left({\kappa M_L\over M_R}\right)^2 2M^2_{Z_1}\sum_i {\tilde C_i
(v^R_i+a^R_i)F_i} \right. \nonumber\\
& + & \left.  2N_\nu M^4_{Z_1}\sum_{i,j}(v^R_iv^R_j+a^R_ia^R_j)
(C_iC_j+\tilde C_i\tilde C_j)P_{ij}\right] \nonumber
\end{eqnarray}
Here, $N_\nu=3$ represents the number of light neutrino species,
\begin{eqnarray}
F_i & = & {M^2-M^2_{Z_i}\over D_i} \,,\nonumber \\
D_i & = & (M^2-M^2_{Z_i})^2+(\Gamma_iM_{Z_i})^2 \,,\\
P_{ij} & = &{(M^2-M^2_{Z_i})(M^2-M^2_{Z_j})+(\Gamma_iM_{Z_i})(\Gamma_iM_{Z_j})
\over  D_iD_j} \,.\nonumber
\end{eqnarray}
The polarized couplings of the electron to the $Z_i$ are related to the
unpolarized couplings $v_i$ and $a_i$ by
\begin{eqnarray}
v_i^{L,R} & = & {1\over 2}(v_i+\lambda a_i)\,, \nonumber\\
a_i^{L,R} & = & {1\over 2}(a_i+\lambda v_i)\,,
\end{eqnarray}
with $\lambda=+1(-1)$ for left-(right-)handed electrons.
$C_i$ and $\tilde C_i$ represent the couplings
of $\nu_L$ and $\nu_R$, respectively, to the $Z_i$.  (Note that $\tilde C_1=0$.)
The couplings are normalized as
\begin{equation}
{\cal L}={g\over 2c_w}\bar e\gamma_\mu(v_i-a_i\gamma_5)eZ_i^\mu \,.
\end{equation}
In evaluating these subprocess cross sections we neglect mixing between the
SM and heavy gauge bosons.

The corresponding subprocess cross sections in the UUM are
\begin{eqnarray}
\sigma_L(x) & = & {G_F^2M^2\over 6\pi}\left\{ \left[ 1+\left( {t_\phi M_L\over
M_H}\right)^2\right]^2 \right. \nonumber\\
& - & \left. 2M^2_{Z_1}\left[ 1+\left( {t_\phi M_L\over
M_H}\right)^2\right]\sum_i {C_i(v_i^L+a_i^L)}F_i\right.
\nonumber\\
& + & \left. 2N_\nu M^4_{Z_1}\sum_{i,j}(v_i^Lv_j^L+a_i^La_j^L)C_iC_jP_{ij} 
\right\}\,,\\
\sigma_R(x) & = & {G_F^2M^2\over 6\pi} 2N_\nu M^4_{Z_1}
\sum_{i,j}(v_i^Rv_j^R+a_i^Ra_j^R)C_iC_jP_{ij} \,. \nonumber 
\end{eqnarray}

We first examine the resulting unpolarized $E_\gamma$ distribution.  Figure 1a 
displays this spectrum for the SM, LRM, and UUM, corresponding to the solid, 
dashed, and dotted curves, respectively, for various values of the parameters.  
It is clear from the figure that the
differences between the SM and the extended models is very small; in fact the SM
and the LRM are indistinguishable on this scale.  In order
to better quantify the influence of the new gauge bosons, we present in Fig. 1b
the ratio of the difference between the differential cross section in the 
extended model from that of the SM to the SM, \ie, $(d\sigma - d\sigma_{SM})/
d\sigma_{SM}$.  We see that the effects of the extended
electroweak sector are at most at the few percent level.  Examination of
the $E_\gamma$ distribution with polarized beams yields similar results.

\nn
\begin{figure}[htbp]
\centerline{
\psfig{figure=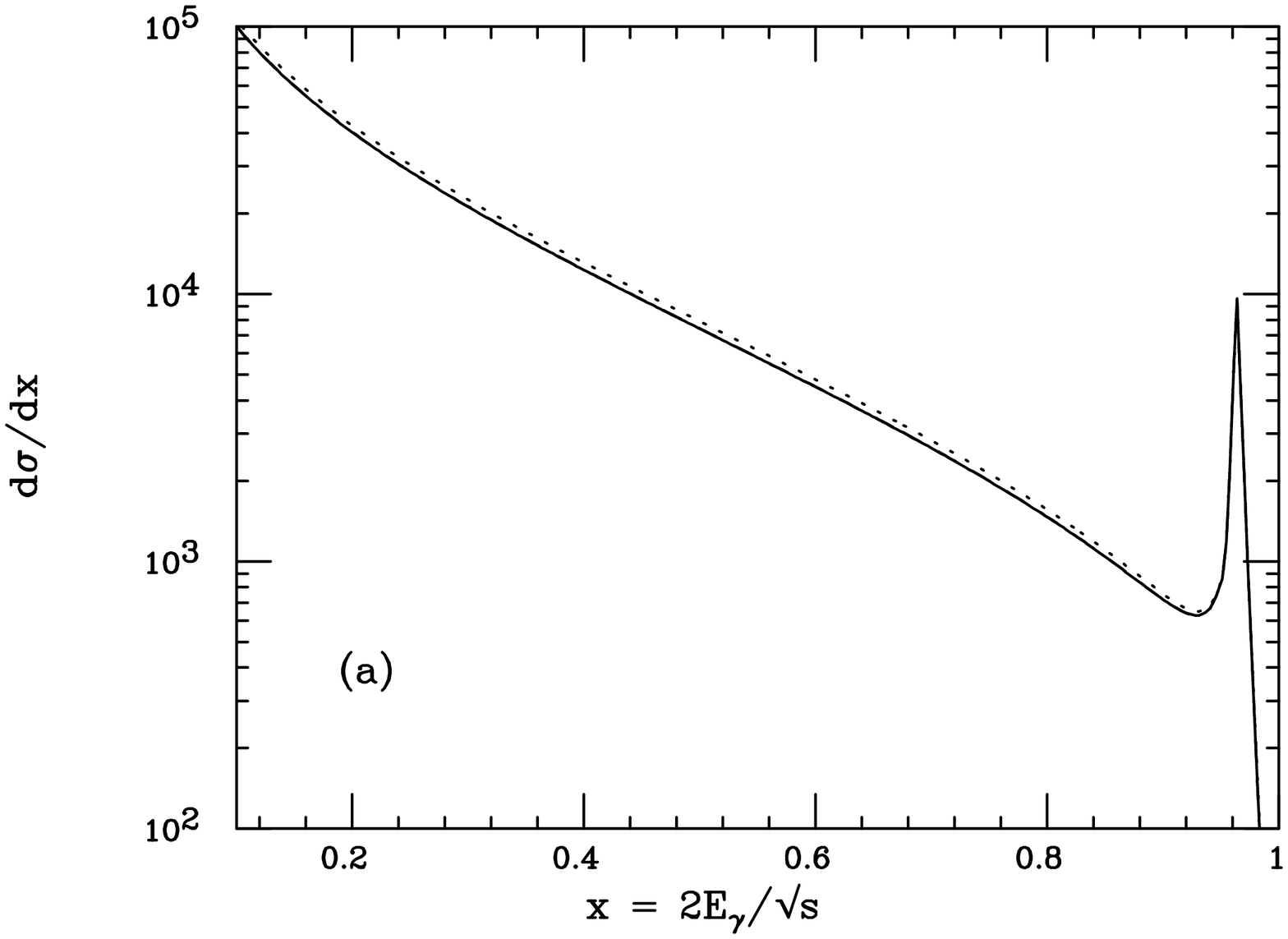,height=6.cm,width=8cm,angle=-90}}
\vspace*{-0.75cm}
\centerline{
\psfig{figure=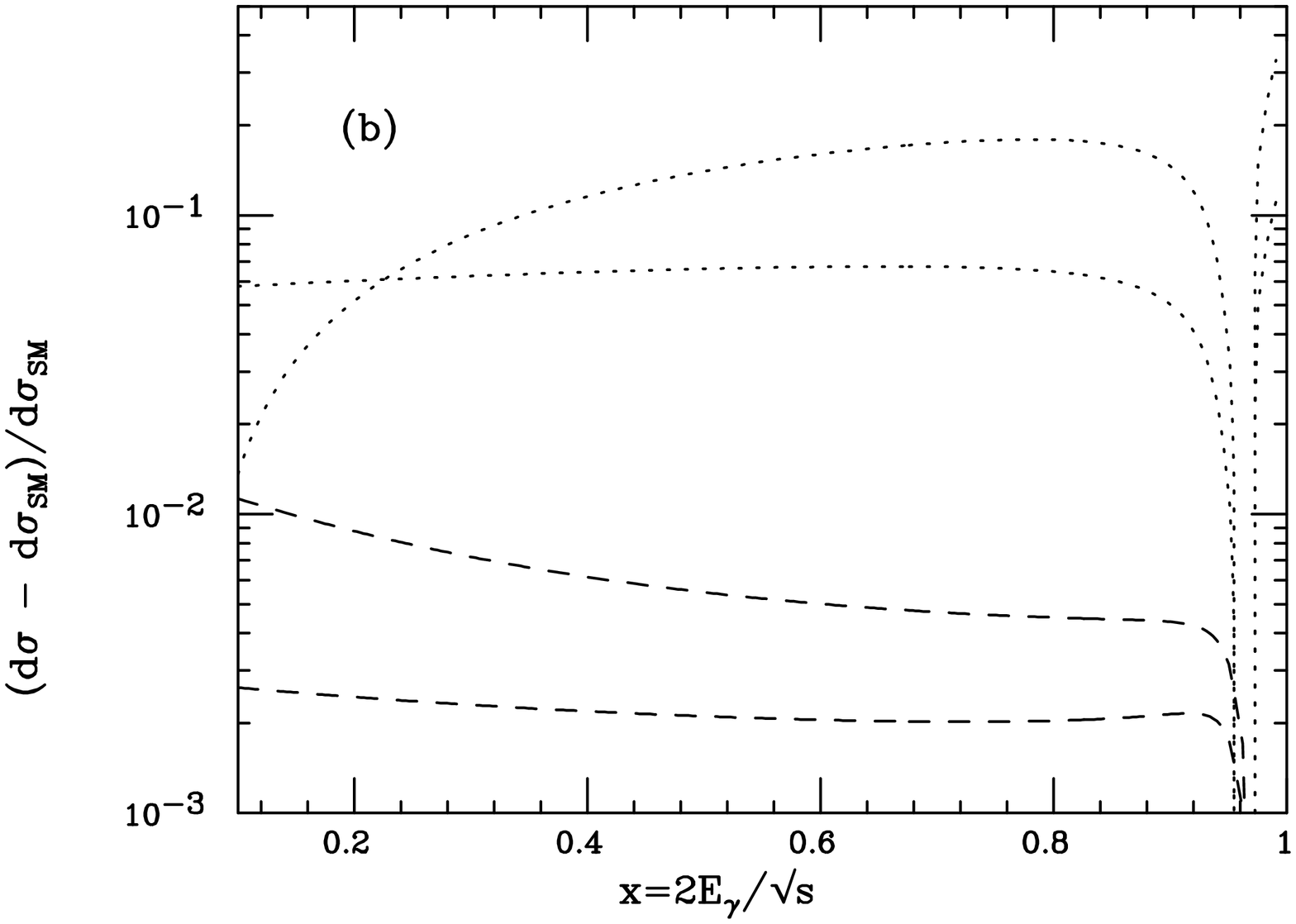,height=6.cm,width=8cm,angle=-90}}
\vspace*{-0.75cm}
\caption{(a) The photon energy spectrum for the SM (solid curve), LRM
with $\kappa=2$ and $M_R=750$ GeV (dashed), and UUM with $\sin\phi=0.6$ and
$M_H=1$ TeV (dotted).  (b) The ratio of the difference of the $E_\gamma$
spectrum in the extended model from the SM to the SM for the LRM with $\kappa=2$
and $M_R=750 (600)$ GeV corresponding to the bottom (top) dashed curves, and
for the UUM with $\sin\phi=0.6$ and $M_H=1 (0.6)$ TeV represented by the
bottom (top) dotted curves.}
\label{edist}
\end{figure}

\begin{figure}[t]
\centerline{
\psfig{figure=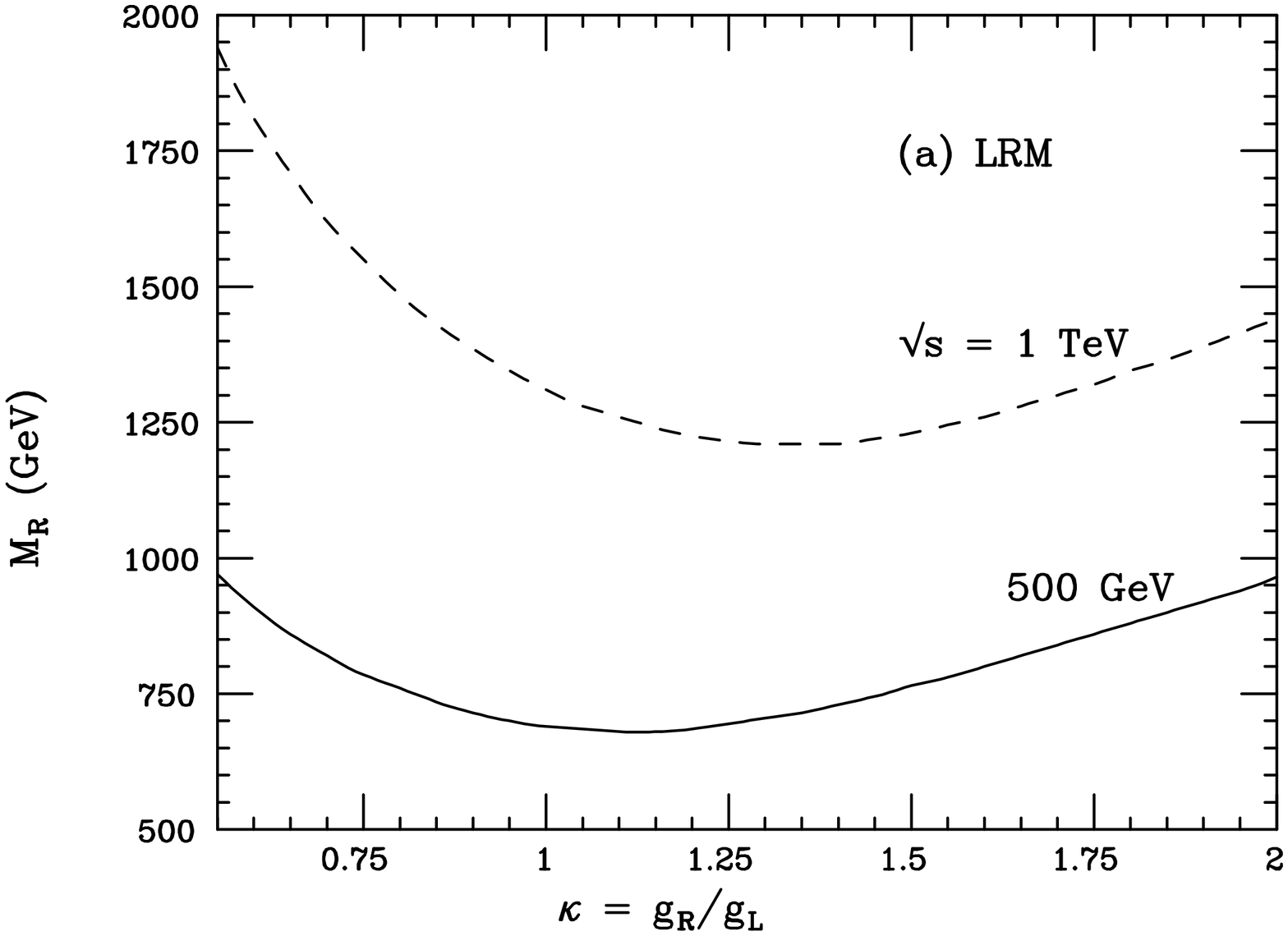,height=6.cm,width=8cm,angle=-90}}
\vspace*{-0.75cm}
\centerline{
\psfig{figure=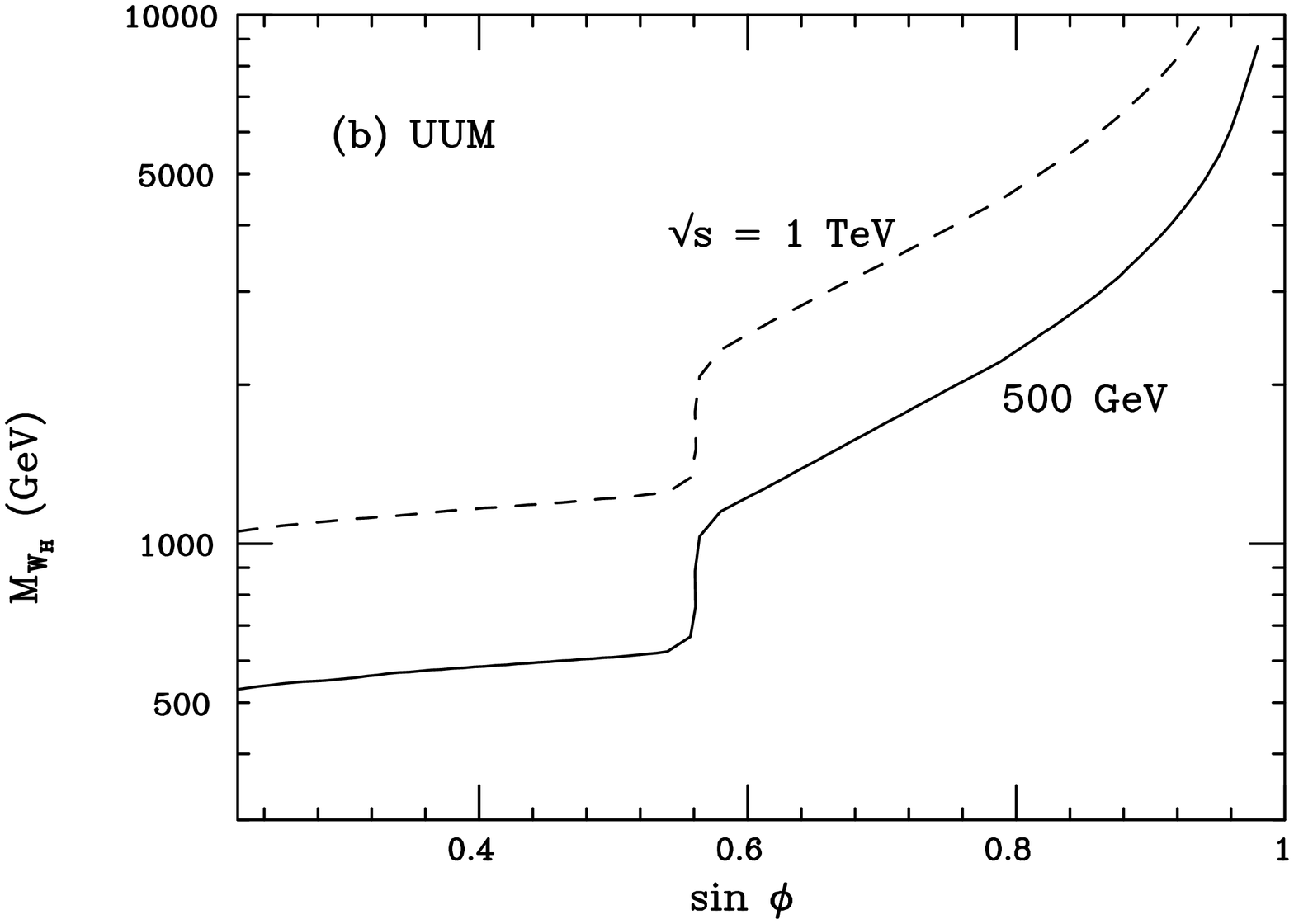,height=6.cm,width=8cm,angle=-90}}
\vspace*{-0.75cm}
\caption{$95\%$ C.L. discovery reach for a new charged gauge boson
via the reaction \nng\ in (a) the LRM as a function of $\kappa$ (b) the UUM 
as a function of $\sin\phi$ for $\sqrt s = 500$ GeV with 50 \infb\ (solid
curve) and 1 TeV with 200 \infb\ (dashed curve).}
\label{bounds}
\end{figure}

In determining the $W'$ search reach we make use of the differential cross
section as well as the left-right asymmetry.
For finite polarization, $P$, these quantities can be written as
\begin{equation}
\sigma(\pm P) = {1\over 2}(1\pm P)\sigma_L+{1\over 2}(1\mp P)\sigma_R \,,
\end{equation}
and 
\begin{equation}
A_{LR} = P {\sigma_L-\sigma_R\over\sigma_L+\sigma_R} \,.
\end{equation}
In our calculations we take the polarization to be $90\%$.
We divide the photon energy spectrum into 10 bins of equal size and perform 
a $\chi^2$ analysis according to usual prescription,
\begin{equation}
\chi^2 = \sum_i \left[{{\cal O}_i - {\cal O}_i^{SM}\over \
\delta{\cal O}_i}\right]^2 \,,
\end{equation}
for our two observables (denoted as ${\cal O}_i$) $d\sigma/dx$ and $A_{LR}$.
We include statistical errors only, such that $\delta\sigma=\sigma/\sqrt{N}$
and $\delta A = \sqrt{(1-P^2A^2)/P^2N}$ for finite polarization, with $N$ being 
the number of events in each bin.  Surprisingly, we find that the left-right
asymmetry contributes very little to the overall value of $\chi^2$.
The resulting $95\%$ C.L. search reach for
heavy charged gauge bosons is presented in Fig. \ref{bounds} for (a) the LRM 
as a function of $\kappa$ and (b) the 
UUM as a function of $\sin\phi$ for $\sqrt s=0.5$ TeV with 50 \infb of 
luminosity and $\sqrt s=1$ TeV with 200 \infb.  In the LRM we see that the
search reach for $W_R$ is expanded to at most $2\times\sqrt s$.  This does
not compete with the discovery potential at the LHC, however it is independent
of assumptions about the right-handed CKM mixing matrix.  In the UUM, the
$W_H$ discovery reach barely extends above $\sqrt s$ for small values of
$\sin\phi$.  However, for larger values of $\sin\phi$ the reach grows
to several times $\sqrt s$ due to the increase in the leptonic couplings for
$\sin\phi>0.5$.
At this stage it is difficult to decouple the effects from the charged gauge 
boson from that of the neutral $Z'$.  
Once the $W$ and $W'$ contact approximation is removed\cite{jlh} we expect that 
appropriate cuts coupled with an examination of the photon angular distribution
will distinguish the $W'$ contribution from that of the $Z'$.  

\nn

In conclusion, we have performed a preliminary study of the reaction \nng\
in two extended gauge models and have found that it does probe the indirect
effects of a heavy charged gauge boson. 
Although the results presented here do not directly compete with the discovery
reach for such particles at the LHC, they do provide proof of demonstration 
that it is possible to observe $W'$ signals with masses is excess of $\sqrt s$
at \epem\ colliders.
This reaction could also serve as a diagnostic tool by providing information 
on the couplings of the new charged gauge boson.


\def\MPL #1 #2 #3 {Mod. Phys. Lett. {\bf#1},\ #2 (#3)}
\def\NPB #1 #2 #3 {Nucl. Phys. {\bf#1},\ #2 (#3)}
\def\PLB #1 #2 #3 {Phys. Lett. {\bf#1},\ #2 (#3)}
\def\PR #1 #2 #3 {Phys. Rep. {\bf#1},\ #2 (#3)}
\def\PRD #1 #2 #3 {Phys. Rev. {\bf#1},\ #2 (#3)}
\def\PRL #1 #2 #3 {Phys. Rev. Lett. {\bf#1},\ #2 (#3)}
\def\RMP #1 #2 #3 {Rev. Mod. Phys. {\bf#1},\ #2 (#3)}
\def\ZPC #1 #2 #3 {Z. Phys. {\bf#1},\ #2 (#3)}
\def\IJMP #1 #2 #3 {Int. J. Mod. Phys, {\bf #1},\ #2 (#3)}


\begin{thebibliography}{99}

\bibitem{tev} 
K. Maeshima, talk presented at {\it 28th International Conference on High
Energy Physics}, Warsaw, Poland, July 1996; S. Eno, \ibid\

\bibitem{future}
T.G. Rizzo, these proceedings, hep-ph/9612440, and 9609248; S. Godfrey, these
proceedings, hep-ph/9612438; M. Cvetic and S. Godfrey, to appear in {\it
Electroweak Symmetry Breaking and Physics Beyond the Standard Model},
ed. T. Barklow \etal.; J.L. Hewett in the {\it Proceedings of the 10th
Topical Workshop on Proton-Antiproton Collider Physics}, Batavia, IL, May
1995, hep-ph/9507400.

\bibitem{d0wp}
S. Abachi \etal, (D0 Collaboration), \PRL 76 3271 1996 .

\bibitem{jlhtgr}
J.L. Hewett and T.G. Rizzo, \IJMP A4 4551 1989 , and in {\it Proceedings of
the DPF Summer Study on High Energy Physics in the 1990's}, Snowmass, CO,
July 1988, ed. by S. Jensen (World Scientific, Singapore 1989); A. Djouadi
\etal, \ZPC C56 289 1992 .

\bibitem{tgr1}
T.G. Rizzo, \PRD D50 5602 1994 .

\bibitem{nlcstudy}
S. Kulhman \etal, (NLC Accelerator Design and Physics Working Group), {\it
Physics and Technology of the Next Linear Collider}, BNL-52502, hep-ex/9605011.

\bibitem{nngam}
T.G. Rizzo, \PRD D34 3516 1986 ; V. Barger, N.G. Deshpande, and K. Whisnant,
\PRL 57 2109 1986 ; V.D. Angelopoulos, \PLB B180 353 1986 .

\bibitem{lrm}
For a review and original references, see, R.N. Mohapatra, {\it Unification
and Supersymmetry}, (Springer, New York, 1986).

\bibitem{desh}
N.G. Deshpande, E. Keith, and T.G. Rizzo, \PRL 70 3189 1993 .

\bibitem{tgr2}
T.G. Rizzo, \PRD D50 325 1994 .

\bibitem{uum}
H. Georgi, E.E. Jenkins, and E.H. Simmons, \PRL 62 2789 1989 , and
\NPB B331 541 1990 ; V. Barger and T.G. Rizzo, \PRD D41 946 1990 ; T.G.
Rizzo, \IJMP A7 91 1992 ; R.S. Chivukula, E.H. Simmons, and J. Terning,
\PLB B346 284 1995 .

\bibitem{approx}
M. Igarashi, N. Nakazawa, and K. Tobimatsu, Prog. Theor. Phys. {\bf 82},
1133 (1989).

\bibitem{jlh}
J.L. Hewett, in preparation.

\end{thebibliography}
\end{document}